# Collapse of CuO Double Chains and Suppression of Superconductivity in High-Pressure Phase of YBa$_2$Cu$_4$O$_8$

Atsuko Nakayama[1]*, Yusuke Onda[2], Shuhei Yamada[2], Hiroshi Fujihisa[3], Masafumi Sakata[4], Yuki Nakamoto[4], Katsuya Shimizu[4], Satoshi Nakano[5], Ayako Ohmura[1], Fumihiro Ishikawa[6], and Yuh Yamada[6]

[1]*Center for Transdisciplinary Research, Niigata University, Niigata 950-2181, Japan*
[2]*Graduate School of Science and Technology, Niigata University, Niigata 950-2181, Japan*
[3]*Research Institute of Instrumentation Frontier, National Institute of Advanced Industrial Science and Technology (AIST), Tsukuba, Ibaraki 305-8565, Japan*
[4]*Center for Science and Technology under Extreme Conditions, Osaka University, Toyonaka-shi, Osaka 560-8531, Japan*
[5]*National Institute for Materials Science, Tsukuba, Ibaraki 305-0044, Japan*
[6]*Department of Physics, Faculty of Science, Niigata University, Niigata 950-2181, Japan*

The crystal structure and electrical resistivity of YBa$_2$Cu$_4$O$_8$ (Y124) were studied under high pressure up to 18 GPa using diamond-anvil cells, respectively, in order to clarify its conduction mechanism. Y124 causes the first-order phase-transition into the orthorhombic *Immm* at pressure around 11 GPa. The high-pressure phase (HPP) also shows the superconductivity, while the manner of temperature dependence of electrical resistance and the pressure dependence of transition temperature, $T_c$, drastically change above 11 GPa. The CuO$_2$ plane persists in HPP but the CuO double chains collapse with the phase transition and transform into three-dimensional Cu-O network, resulting in the renewal of conduction system.

Previous high-pressure experiments have provided us with valuable information to help us find superconductors with higher values of transition temperature, $T_c$. For example the value of $T_c$ of $HgBa_2Ca_2Cu_3O_8$, which is 134 K at atmospheric pressure, increases with increasing the pressure, resulting in 164 K at 31 GPa.[1, 2] $YBa_2Cu_4O_8$ (Y124) is also one of the high-$T_c$ superconductors with $T_c$ = 80 K at atmospheric pressure.[3] The pressure dependence of $T_c$ has a linear behavior up to 5 GPa; while the value of $T_c$ decreases after passing through the maximum $T_c$ = 108 K at 12 GPa.[4] Ludwig et al. observed continues changes in lattice parameters of Y124 under pressure up to 25 GPa using a diamond anvil cell (DAC) and NaF pressure-transmitting medium (PTM).[5] Sholtz et al. measured the resistance of bulk ceramic of Y124 in magnetic fields up to 10 T, pressures to 18 GPa, and temperature above 50 K, resulting in the parabolic and continuous pressure dependence on $T_c$.[6]

The distance from the $CuO_2$ plane to the oxygen atom located at the top of the tetragonal pyramid, $d_{Cu-O}$, gives important information to help us understand the pressure effect of $T_c$.[7, 8] Nelmes et al., investigated the pressure change in the crystal structure of Y124 up to 5 GPa, and clarified that the coefficient of contraction in $d_{Cu-O}$, is twice of that in the linier compressibility of $c$-axis length. This means that the number of holes on $CuO_2$ plane increases with the increase of pressure.

On the other hand, recent research has pointed out the anomaly of pressure-induced structural-change of Y124 under pressures above 3.7 GPa.[9] Calamiotou et al. have investigated the pressure dependences of lattice constants through synchrotron powder x-ray diffraction (XRD) measurement using a 4:1 methanol-ethanol mixture as PTM. Their research revealed that Y124 shows pressure-induced lattice anomalies and hysteresis in the pressure dependence of the lattice parameters in the range from 3.7 to 12 GPa.

The discrepancy between the two reports on structure (Refs. 5 and 9), should be induced by hydrostaticity from PTM. The use of helium (He) as PTM brings quasi-hydrostaticity under pressure over 10 GPa significantly decreases the effect of uni-axial stress, resulting in the sharp diffraction pattern under high pressure.[10] It gives intrinsic structural information at each pressure point.

In order to clarify the conduction mechanism of Y124 under high pressure especially above 10 GPa, we carried out the following experiments: (1) powder XRD under pressure up to 18 GPa and at room temperature (RT) under hydrostatic condition using He PTM, and (2) temperature dependence of electrical resistance using a single crystal of Y124 by cooling down to 4 K under pressure up to 18 GPa.

The polycrystalline Y124 compound for powder XRD measurement was synthesized by a high-oxygen pressure-technique.[3] The precursor was prepared using a polymerized-complex method of an appropriate amount of $Y_2O_3$, CuO and $Ba(NO_3)_2$, which were decomposed at 500 ºC in air for 24 h. The precursor was pressed into a pellet and reacted in pure oxygen gas of 10 atm at 900 ºC for 48 h. The sample obtained was ground into fine powders in an alumina mortar cooled in liquid nitrogen. We also prepared the single crystal for the electrical resistance measurements, which was grown by cooling of $YBaCu_3O_6$-CuO mixture (1:1 in mole) from 725 ˚C down to 450 ˚C for 4 h, using a KOH flux method.[11]

XRD patterns were measured by using the DAC set up a couple of diamond anvils with 0.6-mm culet-diameters and 2-mm anvil-thicknesses. The gasket was made of 65-µm-thick stainless steel, in which a 260-µm hole was drilled. A small amount of powdered sample was put in the gasket hole mounted on the culet of lower anvil with ruby balls, which was filled with high-density He-gas.[12] Each pressure was determined from the fluorescence peaks of ruby balls put in the sample chamber on the basis of ruby pressure scale.[13]

Angle-dispersive powder-patterns were taken using synchrotron radiation (SR) from the bending magnet on the beam-line BL-18C at Photon Factory, High Energy Accelerator Research Organization (KEK). The SR beam was monochromatized to a wavelength 0.6190 Å and introduced to the specimen through a pinhole collimator (80 µm diameter). Each pattern was obtained by having the sample exposed to the x-ray for 20 min at RT. An imaging plate and a BAS2500 scanner, supplied by Fuji Photo Film Co. Ltd., were used to obtain two-dimensional powder diffraction image. One image has 2000×2560 pixels with a resolution of $100 \times 100$ µm$^2$ per one pixel.

The electrical resistance of Y124 was measured under pressure up to 18 GPa by a four-probe method in DAC. A rhenium metal gasket was insulated from the sample and electrodes using a mixture of cubic-boron nitride and an epoxy resin. Gold foils were connected to the platinum terminals, which were sputtered on the *ab*-plane of crystal with a dimension of 90×50×20 µm$^3$. Sodium chloride was used as PTM. Several ruby chips were put around the sample for the pressure determination. The resistance was measured by an ac method with an excitation current smaller than 100 $\mu$A.

The powder XRD patterns of Y124 obtained under pressure up to 18.1 GPa at RT are shown in Fig. 1(a). The pressurization above 10.1 GPa causes the change in the pattern accompanied with drastic decrease of the 00*l* reflections (*l* = 2, 4 and 6). The 104 and 106 reflections derived from the Cu-O double-change structure disappeared at 11.0 GPa. In view of these results we identified that the structural-phase transition occurs at around 11 GPa. Each reflection peak moved to the lower-angle direction during reducing the pressure; the XRD pattern of released sample finally recovered to the original.

The crystal structure of high-pressure phase (HPP) has been determined by Rietveld

analysis[14]) and molecular dynamic (MD) simulation based on a density-functional theory (DFT). The DFT calculations for structural optimizations and MD simulations were carried out using the program MS CASTEP of Accelrys, Inc.[15]) We employed the generalized-gradient approximation (GGA) - Perdew-Burke- Ernzerhof for solids (PBEsol) exchange correlation functional[16]) and ultrasoft pseudopotentials[17]) with the energy cut-off of 440 eV. The followings are the atomic positions of oxygen estimated by the MD simulation of crystal structure at 11.0 GPa: O1 (0.5, 0.5, 0.3496), O2 (0, 0.5, 0.4402), O3 (0, 0.5, 0.0618) and O4 (0, 0.5, 0.2343). We bound the positions of all oxygen atoms using them in the Rietveld analyses of HPP.

The Rietveld fit shown in Fig. 1(b), revealed that the structure after the phase transition is assigned to be an orthorhombic with the space group *Immm*. The lattice parameters were obtained to be $a = 3.746 \pm 0.001$ Å, $b = 3.833 \pm 0.001$ Å, and $c = 25.324 \pm 0.005$ Å at 11.0 GPa. In this case, the atomic positions of Y, Ba, and Cu were given using xyz coordinates as follows: Y (0.5, 0.5, 0), Ba (0.5, 0.5, 0.1400), Cu1 (0.5, 0.5, 0.2793) and Cu2 (0.5, 0.5, 0.4389). Wang *et al*. also reported the phase transition to orthorhombic at 11 GPa; they chose non-centrosymmetric orthorhombic the space group *Imm*2.[18]) Both *Immm* and *Imm*2 space groups are explained by a same extinction rule; either is suitable to explain the peak positions observed in the XRD patterns of HPP. Analyses of internal structures using both space groups, respectively, produced the same structure. Therefore we chose the centrosymmetric space group *Immm* for the crystal structure of HPP.

The pressure changes in lattice parameters *a*, *b*, and *c*, and volume *V*, are shown in Fig. 2(a) and (b), respectively. The lattice parameters obtained at the low-pressure phase (LPP) are determined to be $a = 3.739 \pm 0.001$ Å, $b = 3.825 \pm 0.001$, $c = 26.33 \pm 0.01$ Å at 10.1 GPa, respectively. The *c*-axis length obtained at 11.0 GPa is 3.8 % shorter than that at 10.1 GPa. The *b*-axis length temporary increases just after the phase transition, which shows the decrease with increasing the pressure in HPP. The variation in volumes measured before and after the transition is obtained to be $\Delta V / V = -3.83$ %.

Projections on the *ab* plane of crystal structures obtained before and after the phase transition are shown in Fig. 3. The phase transition of Y124 is explained as follows: The unit cell is pulled in opposite directions along the *a* axis, resulting in deformation into the monoclinic in HPP. (See the gray colored cells.) We were able to replace the monoclinic with the orthorhombic, which is higher in the symmetry than the monoclinic. In this way the crystal structure of HPP was determined to be the *Immm* structure. Every quadrangular pyramid displayed in green color, consisting of the apical oxygen atom and $CuO_2$ basal plane, maintains after phase transition. However the CuO double-chains structure, perpendicular to the *ac* plane, is broken after phase transition and three-dimensional Cu-O network are generated in HPP.

Temperature dependence of electrical resistance measured under pressure up to 17.2 GPa is shown in Fig. 4(a). Sharp falls of the resistance based on the superconducting transition were observed at the pressures up to 9.7 GPa. Pressurization exceeding 9.7 GPa, caused dramatic changes in the temperature dependences; the resistance is depressed in the temperature range which is higher than $T_c$, and indicates the broad temperature dependence observed in the process to reach the zero resistance. As shown in Fig. 4 (b), the values of both $T_{c\text{-onset}}$ and $T_{c\text{-zero}}$ drastically decrease at above 9.7 GPa. That is clearly different from the previous result.[6] According to the recent report, the Meissner effect in Y124 was found to disappear under the uniaxial strain above 10 GPa.[20]

The onset of superconductivity above 12.4 GPa does not occur in two steps, meaning the single phase in HPP. The broad transition may be brought by distribution of $T_c$ based on the pressure gradient in the sample. While NaCl is less efficient as the PTM than He, it is enough to obtain the resistance data in the pressure range. Assuming that the pressure-induced structural-phase transition occurs at around 11 GPa at low temperature, the negative pressure dependence of both $T_{c\text{-onset}}$ and $T_{c\text{-zero}}$ shown in Fig. 4 (b) is an essential feature for HPP. It is believed that the depression of $T_c$ observed above 9.7 GPa is caused by collapse of the CuO double chains to supply the carriers. However the $CuO_2$ plane persists in HPP. This means that the $CuO_2$ plane plays the role of conductive layer in both LPP and HPP. After the dramatic decrease of $T_c$ a distribution of conduction electrons on the $CuO_2$ plane seems to change after the phase transition.

In summary both XRD and the electrical-resistance measurements of $YBa_2Cu_4O_8$ show the presence of pressure-induced phase-transition at pressure around 11 GPa. The $CuO_2$ plane persists in HPP, having the orthorhombic *Immm*. The $CuO_2$ plane plays the role of conductive layer in both LPP and HPP. On the other hand the CuO double chains collapse after phase transition, resulting in another Cu-O bonding structure. Not only the LPP but also HPP shows the superconductivity; the value of $T_c$ decreases as a result of increased pressures, which are caused by collapsing of the CuO double chains.

The authors thank Dr. Kikegawa of KEK for his help on the BL-18C. This work was carried out under Proposal No. 2009G624 of photon Factory. This work was supported by a Grant-in-Aid for Scientific Research from the Ministry of Education, Science and Culture, Japan.

*E-mail: nakayama-a@phys.sc.niigata-u.ac.jp

1) L. Gao, Y. Y. Xue, F. Chen, Q. Xiong, R. L. Meng, D. Ramirez, C. W. Chu, J. H. Eggert, and H. K. Mao, Phys. Rev. B **50**, 4260 (1994).
2) N. Takeshita, A. Yamamoto, A. Iyo, and H. Eisaki, J. Phys. Soc. Jpn. **82**, 023711 (2013).
3) J. Karpinski, E. Kaldis, E. Jilek, S. Rusiecki, B. Bucher, Nature **336**, 660 (1988).
4) E. N. Van Eenige, R. Griessen, and R. J. Wijngaarden, Physica C **168**, 482 (1990).
5) H. A. Ludwig, W. H. Fietz, M. R. Dietrich, H. Wühl, J. Karpinski, E. Kaldis, and S. Rusiecki, Physica C **167**, 335 (1990).
6) J. J. Scholtz, E. N. van Eenige, R. J. Wijngaarden, and R. Griessen, Phys. Rev. B **45**, 3077 (1992).
7) R. J. Nelmes and J. S. Loveday, Physica C **172**, 311 (1990).
8) Y. Yamada, J. D. Jorgensen, Shiyou Pei, P. Lightfoot, Y. Kodama, T. Matsumoto, F. Izumi, Physica C **173**, 185 (1991).
9) M. Calamiotou, A. Gantis, E. Siranidi, D. Lampakis, J. Karpinski, E. Liarokapis, Phys. Rev. B **80**, 214517 (2009).
10) K. Takemura, Phys. Rev. B **60**, 6171 (1999).
11) G L Sun, Y T Song and C T Lin, Supercond. Sci. Technol. **21** 125001 (2008)
12) K. Takemura, P. Ch, Sahu, Y. Kunii, and Y. Toma, Rev, Sci. Instrum. **72**, 3873 (2001).
13) C. S. Zha, H. K. Mao, and R J. Hemlsy, Proc. Natl., Acad. Sci., **97**, 13494 (2000).
14) F. Izumi, in *The Rietveld Analysis*, ed. R. A. Young, (Oxford University Press, New York, 1993) p. 236.
15) S. J. Clark, M. D. Segall, C. J. Pickard, P. J. Hasnip, M. I. J. Probert, K. Refson, and M. C. Payne, Z. Kristallogr. **220**, 567 (2005).
16) J. P. Perdew, A. Ruzsinszky, G. I. Csonka, O. A. Vydrov, G. E. Scuseria, L. A. Constantin, X. Zhou, and K. Burke, Phys. Rev. Lett. **100**, 136406 (2008).
17) D. Vanderbilt, Phys. Rev. B **41**, 7892 (1990).
18) X. Wang, F. H. Su, S. Karmakar, K. Syassen, Y. T. Song and C. T. Lin, in Max-Planck-Institut für Festkörperforschung Stuttgart Annual Rep., 70 (2007).
19) F. D. Murnaghan, Proceedings of the National Academy of Science of the United States of America **30**, 244 (1944).
20) M. Mito, T. Imakyurei, H. Deguchi, K. Matsumoto, H. Hara, T. Ozaki, H. Takeya and Y. Takano, J. Phys. Soc. Jpn. **83**, 023705 (2014).

Fig. 1. (Color online) (a) Powder XRD patterns of Y124 observed at the pressures up to 18.1 GPa and at RT. The XRD patterns obtained under pressure up to 10.1 GPa are explained with an orthorhombic *Ammm*. (b) Result of Rietveld analysis obtained at 12.2 GPa and at RT. The diffraction patterns were explained using a two-phase model: orthorhombic *Immm* for HPP (pink) and orthorhombic *Ammm* for the LPP (green), respectively. The lattice parameters of *Ammm* structure remaining in the high-pressure phase were predicted by extrapolation from the linear equation of state estimated in the low-pressure phase. (See Fig. 2) In the Rietveld analysis the lattice parameters of *Immm* were obtained by fixing the *Ammm* structure. The volume fraction of *Immm* structure is estimated to be 91.3 % at 12.2 GPa. The indexes for *Immm* symmetry in Fig. 1(a) were determined by the Rietveld fit.

Fig. 2. (Color online) Pressure changes in (a) the lattice parameters *a*, *b*, and *c*, and (b) volume $V$ up to 18.1 GPa at room temperature. By fitting the variation in $V$ with the Murnaghan-type equation of state[19] as shown in (b), the values of bulk modulus $B_0$ and its pressure derivative $B_0'$ for the low pressure phase are determined to be $B_0$ = 108.44 GPa and $B_0'$ = 7.141 with $V_0$ = 405.26 Å$^3$, respectively. The lattice parameters of *Ammm* structure remaining in the high-pressure phase are also plotted in the graphs. Each parameter in *Ammm* structure was predicted by extrapolation from the linear equation of state estimated in the low-pressure phase: $P = (B_0/B_0')\{(x_0/x)^{B_0'}-1\}$, $x = a, b$, and $c$. The values of $B_0$ and $B_0'$ are $B_0$ = 306.94 GPa and $B_0'$ = 17.17 for *a*-axis length with $a_0$ = 3.8443 Å, $B_0$ = 535.46 GPa and $B_0'$ = 60.98 for *b*-axis length with $b_0$ = 3.8723 Å, $B_0$ = 242.85 GPa and $B_0'$ = 13.94 for *c*-axis length with $c_0$ = 27.224 Å, respectively.

Fig. 3. (Color online) Projections on the *ab* plane of crystal structures obtained before and after the phase transition: LPP with *Ammm* structure (left hand) and HPP with *Immm* structure (right hand).

Fig. 4. (Color online) (a) Temperature dependence of electrical resistance under pressure up to 17.2 GPa. (b) Pressure dependence of $T_{\text{c-onset}}$ and $T_{\text{c-zero}}$. The data measured by Scholtz[6] are also shown in the same graph. The dotted curves are drawn for easy visual guidance. The space groups, *Ammm* and *Immm*, were determined at RT, respectively. The insert shows definition of $T_{\text{c-onset}}$ used in this study. We drew two tangent lines in the normal state parts of the resistance-temperature curve before and after the transition. The value of $T_{\text{c-onset}}$ is estimated from the position of intersection point between the two lines.

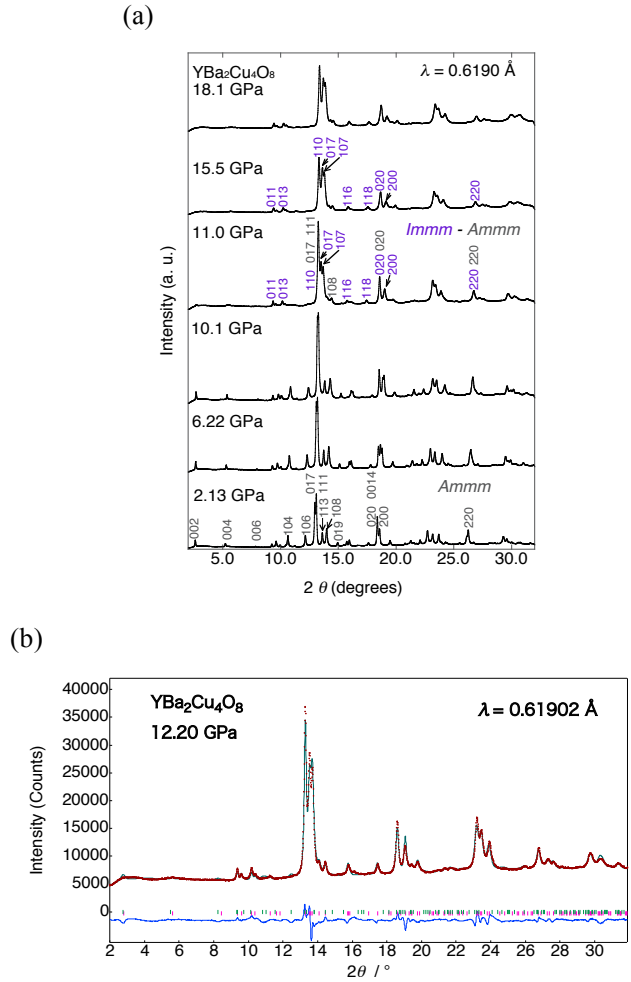

Fig. 1. (Color online) (a) Powder XRD patterns of Y124 observed at the pressures up to 18.1 GPa and at RT. The XRD patterns obtained under pressure up to 10.1 GPa are explained with an orthorhombic *Ammm*. (b) Result of Rietveld analysis obtained at 12.2 GPa and at RT. The diffraction patterns were explained using a two-phase model: orthorhombic *Immm* for HPP (pink) and orthorhombic *Ammm* for the LPP (green), respectively. The lattice parameters of *Ammm* structure remaining in the high-pressure phase were predicted by extrapolation from the linear equation of state estimated in the low-pressure phase (See Fig. 2) In the Rietveld analysis the lattice parameters of *Immm* were obtained by fixing the *Ammm* structure. The volume fraction of *Immm* structure is estimated to be 91.3 % at 12.2 GPa. The indexes for *Immm* symmetry in Fig. 1(a) were determined by the Rietveld fit.

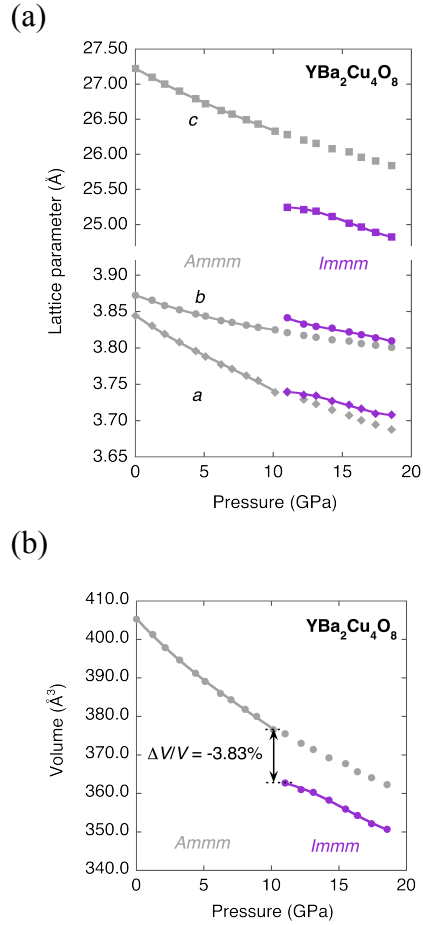

Fig. 2. (Color online) Pressure changes in (a) the lattice parameters *a*, *b*, and *c*, and (b) volume *V* up to 18.1 GPa at room temperature. By fitting the variation in *V* with the Murnaghan-type equation of state[19] as shown in (b), the values of bulk modulus $B_0$ and its pressure derivative $B_0$' for the low pressure phase are determined to be $B_0$ = 108.44 GPa and $B_0$' = 7.141 with $V_0$ = 405.26 Å$^3$, respectively. The lattice parameters of *Ammm* structure remaining in the high-pressure phase are also plotted in the graphs. Each parameters was predicted by extrapolation from the linear equation of state estimated in the low-pressure phase: $P = (B_0/ B_0')\{(x_0/x)^{B_0'}-1\}$, $x =$ *a*, *b*, and *c*. The values of $B_0$ and $B_0$' are $B_0$ = 306.94 GPa and $B_0$' = 17.17 for *a*-axis length with $a_0$ = 3.8443 Å, $B_0$ = 535.46 GPa and $B_0$' = 60.98 for *b*-axis length with $b_0$ = 3.8723 Å, $B_0$ = 242.85 GPa and $B_0$' = 13.94 for *c*-axis length with $c_0$ = 27.224 Å, respectively.

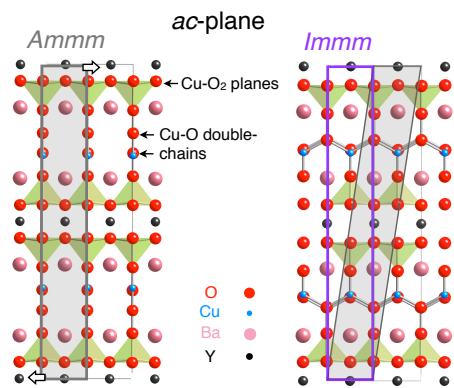

Fig. 3. (Color online) Projections on the *ab* plane of crystal structures obtained before and after the phase transition: LPP with *Ammm* structure (left hand) and HPP with *Immm* structure (right hand).

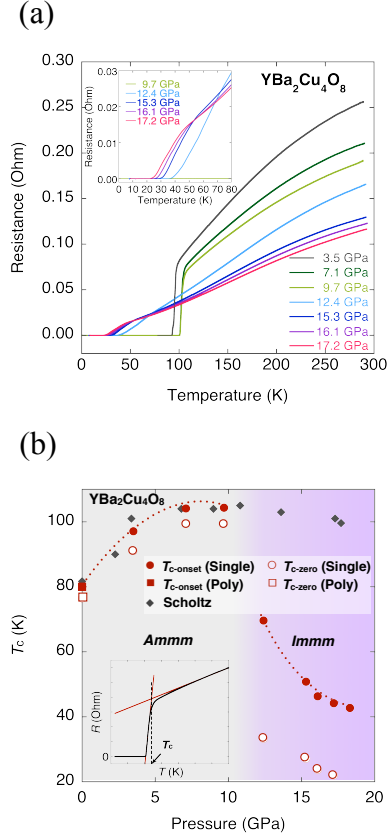

Fig. 4. (Color online) (a) Temperature dependence of electrical resistance under pressure up to 17.2 GPa. (b) Pressure dependence of $T_{c\text{-onset}}$ and $T_{c\text{-zero}}$. The data measured by Scholtz[6] are also shown in the same graph. The dotted curves are drawn for easy visual guidance. The space groups, *Ammm* and *Immm*, were determined at RT, respectively. The insert shows definition of $T_{c\text{-onset}}$ used in this study. We drew two tangent lines in the normal state parts of the resistance-temperature curve before and after the transition. The value of $T_{c\text{-onset}}$ is estimated from the position of intersection point between the two lines.